\documentclass[final,3p,times,twocolumn]{elsarticle}
 \biboptions{comma,sort&compress}
\usepackage{here}
 \usepackage{graphicx}
  \usepackage{epsfig}
  \usepackage{slashed}
  \usepackage{url}
\def\nin{\noindent}
\def\beq{\begin{equation}}
\def\eeq{\end{equation}}
\def\bea{\begin{eqnarray}}
\def\eea{\end{eqnarray}}

\journal{Nuc. Phys. (Proc. Suppl.)}

\begin{document}

\begin{frontmatter}



\title{Glueball spectrum and hadronic processes in low-energy QCD}

 \author[label1]{Marco Frasca\corref{cor1}}
  \address[label1]{Via Erasmo Gattamelata, 3 \\
  00176 Roma (Italy)}
\cortext[cor1]{Speaker}
\ead{marcofrasca@mclink.it}

%\author{}

%\address{}

\begin{abstract}
\noindent
Low-energy limit of quantum chromodynamics (QCD) is obtained using a mapping theorem recently proved. This theorem states that, classically, solutions of a massless quartic scalar field theory are approximate solutions of Yang-Mills equations in the limit of the gauge coupling going to infinity. Low-energy QCD is described by a Yukawa theory further reducible to a Nambu-Jona-Lasinio model. At the leading order one can compute glue-quark interactions and one is able to calculate the properties of the $\sigma$ and $\eta-\eta'$ mesons. Finally, it is seen that all the physics of strong interactions, both in the infrared and ultraviolet limit, is described by a single constant $\Lambda$ arising in the ultraviolet by dimensional transmutation and in the infrared as an integration constant.
\end{abstract}


\end{frontmatter}

% \linenumbers

\section{Introduction}
%\label{}
\nin
%%%%%%%%%%%%

Understanding low-energy QCD is a key to uncover the structure of light unflavored mesons. To paraphrase Luciano Maiani:''If these are all tetraquark states, where are exotic states?''\cite{lmai}. We will give a first answer to this question.

KLOE-2 measurements\cite{kloe} seem to rule out quark contributions to the structure of lighter mesons. Similarly, their results support a glue component in $\eta'$ meson that, having a glue component in the structure, is seen to possibly emit a glue state that finally decays in two pions. Consistency is only obtained if the emitted glue state is a $\sigma$ meson.

Structure of $\sigma$ resonance is hotly debated. Common view is that this meson should be a tetraquark state, member of a low-lying nonet. This state has not been seen yet in lattice computations either quenched or not. Recent analysis on $\gamma\gamma$ decay and data from NA48/2\cite{nar1} seem to support the idea that this is a glue state rather than a quark composite particle.

Similarly, f0(980) appears to be a possible glue state and can be seen as an excited state of $\sigma$. Mass of this state should take into account its KK decay.

All properties of hadronic processes must be consistent with a single parameter of the theory, $\Lambda$, representing a constant arising in ultraviolet limit by dimensional transmutation and in the infrared limit as an integration constant of the theory. Theory must be working with a single $\Lambda$ parameter, even if some authors admit a possible dependency on the energy scale \cite{bou}. Our aim here is to fix this constant from a low-energy theoretical analysis.

%%%%%%%%%%%%
\section{Mapping theorem and low-energy QCD}
%\label{}
\nin
%%%%%%%%%%%%

In order to manage QCD, it appears essential to find a way to reduce this theory to a simpler one. With this aim in mind the following theorem has been proved:

{\bf MAPPING THEOREM:} {\sl An extremum of the action
\begin{equation}
\nonumber
    S = \int d^4x\left[\frac{1}{2}(\partial\phi)^2-\frac{\lambda}{4}\phi^4\right]
\end{equation}
is also an extremum of the SU(N) Yang-Mills Lagrangian when one properly chooses $A_\mu^a$ with some components being zero and all others being equal, and $\lambda=Ng^2$, being $g$ the coupling constant of the Yang-Mills field, when only time dependence is retained. In the most general case the following mapping holds
\begin{equation}
\nonumber
    A_\mu^a(x)=\eta_\mu^a\phi(x)+O(1/\sqrt{N}g)
\end{equation}
being $\eta_\mu^a$ constant, that becomes exact for the Lorenz gauge.}

A first proof of this theorem was given in \cite{fra1} and, after a criticism by Terence Tao, a final proof was presented in \cite{fra2} also agreed with Tao\cite{tao}.

Applying the above mapping theorem, holding in the large coupling limit, QCD generating functional becomes:
\begin{eqnarray*}
\scriptstyle{
Z[\eta,\bar\eta,j]=\int \prod_q[dq][d\bar q][d\phi]
e^{i(N^2-1)\int d^4x\left[\frac{1}{2}(\partial\phi)^2-\frac{Ng^2}{4}\phi^4\right]}\times}&& \\
\scriptstyle{
e^{i\int d^4x\sum_q\bar q(x)\left[\gamma\cdot\left(i\partial-g\frac{\lambda\cdot\eta}{2}\phi\right)-m_q\right]q(x)}
e^{i\int d^4x\sum_q[\bar q(x)\eta_q(x)+\bar\eta_q(x)q(x)]}\times} && \\
\scriptstyle{e^{i\int d^4x j(x)\cdot\eta\phi(x)}+O(1/\sqrt{N}g).}&&
\end{eqnarray*}
At this order ghost field just decouples so we can safely ignore it. QCD is so reduced to a Yukawa model by the use of the mapping theorem, at the leading order of a development in the inverse of the 't Hooft coupling. All the parameters of the model are fixed by QCD.
In the limit $\lambda\rightarrow\infty$, the scalar field term takes a Gaussian form \cite{fra3}. This theory is trivial in this limit and the beta function goes like $\beta(\lambda)=4\lambda$ in four dimensions at lower momenta.

The generating functional for the Yukawa model can be cast into a Gaussian form in the strong coupling limit $g\rightarrow\infty$, knowing the gluon propagator. Indeed, one can write
\begin{eqnarray*}
\scriptstyle{
Z[\eta,\bar\eta,j_\phi]\approx
    e^{\frac{i}{2}(N^2-1)\int d^4xd^4y j_\phi(x)\Delta(x-y)j_\phi(y)}\int\prod_q[dq][d\bar q]\times}&& \\
    \scriptstyle{
    e^{i\int d^4x\sum_q\bar q(x)\left[i\gamma\cdot\partial+g\frac{\lambda^a}{2}
    \gamma\cdot\eta^a\int d^4y\Delta(x-y)j_\phi(y)\right]q(x)}\times}&& \\
    \scriptstyle{
    e^{i\frac{1}{2}\frac{g^2}{4}\int d^4xd^4y\sum_{q,q'}\bar q(x)\lambda^a\eta^a_\mu\gamma^\mu q(x)
      \Delta(x-y)\bar q'(y)\lambda^b\eta^b_\nu\gamma^\nu q'(y)}\times}&& \\
      \scriptstyle{
    e^{i\int d^4x\sum_q[\bar q(x)\eta_q(x)+\bar\eta_q(x)q(x)]}}&&.
\end{eqnarray*}
provided $j_\phi=j\cdot\eta$ and the gluon propagator
%and $\partial_t^2G(t_x-t_y)+Ng^2[G(t_x-t_y)]^3=\delta(t_x-t_y)$ and so
\begin{equation}
\nonumber
    \Delta(p)=\sum_{n=0}^\infty\frac{B_n}{p^2-m_n^2+i\epsilon}
\end{equation}
being
\begin{equation}
\nonumber
    B_n=(2n+1)\frac{\pi^2}{K^2(i)}\frac{(-1)^{n+1}e^{-(n+\frac{1}{2})\pi}}{1+e^{-(2n+1)\pi}}.
\end{equation}
We get a first key formula and this is the spectrum of the theory, in the strong coupling limit, given by
\begin{equation}
\label{eq:spe}
    m_n = \left(n+\frac{1}{2}\right)\frac{\pi}{K(i)}\left(\frac{Ng^2}{2}\right)^{\frac{1}{4}}\Lambda.
\end{equation}
From the mass spectrum we can identify a string tension as
\begin{equation}
\nonumber
    \sqrt{\sigma}=\left(\frac{Ng^2}{2}\right)^{\frac{1}{4}}\Lambda=(2\pi N\alpha_s)^{\frac{1}{4}}\Lambda.
\end{equation}
Presently, the parameter $\Lambda$ appears rather arbitrary. Being an integration constant, it should be obtained from experiment.
We just note from this that $\sigma_{SU(2)}/\sigma_{SU(3)}=\sqrt{2/3}$ as seen on lattice\cite{tep1}.
We recognize that, at lower energies, strong interactions are mediated by a kind of bosons that can be seen as due to Yang-Mills field self-interaction. These are the physical states in a strong coupling limit.

We realize that the low-energy limit of QCD can be further reduced to a Nambu-Jona-Lasinio model. In the gluon propagator we just take the low momenta limit producing the contact interaction\cite{fra4}
\begin{equation}
\nonumber
    \Delta(x-y)\approx \frac{3.76}{\sigma}\delta^4(x-y).
\end{equation}
A first analysis of simple interactions may be accomplished by neglecting quark loops or decays involving quarks, just quark-glue vertexes. So, for a first understanding we just consider quark-glue interaction.
In this case, QCD is exactly integrable producing a non-trivial Gaussian generating functional.

So, the generating functional can be finally integrated producing the final result
%\small
\begin{eqnarray*}
    &&\scriptstyle{Z[\eta,\bar\eta,j_\phi]\approx
    \exp\left\{\frac{i}{2}(N^2-1)\int d^4xd^4y j_\phi(x)\Delta(x-y)j_\phi(y)\right\}\times} \\
    &&\scriptstyle{\exp\left\{i\int d^4xd^4y\sum_q\bar\eta_q(x)S[j_\phi,x-y]\eta_q(y)\right\}}
\end{eqnarray*}
%\normalsize
The quark propagator, considering that {\sl a gradient expansion corresponds to a strong coupling expansion}\cite{fra7}, is
\begin{eqnarray*}
    &&S[j_\phi,x-y]=\theta(t_x-t_y)\delta^3(x-y)\times \\
    & &e^{\left\{ig\frac{\lambda^a}{2}
    \gamma_0\gamma_i\eta_i^a\int_{t_x}^{t_y} dt'
    \int d^4x_1 \Delta(t'-t_y-t_{x_1},x-y-x_1)j_\phi(x_1)\right\}}.
\end{eqnarray*}
We have a defined leading order term for a strong coupling expansion in QCD. Higher order terms contain gradients of the fields.
QCD at the leading order in a strong coupling expansion ($g\rightarrow\infty$) appears a confining and yet renormalizable theory.

\section{$\sigma$ and $\eta-\eta'$ mesons}

We can identify the $\sigma$ meson as the lowest state in Yang-Mills theory. Being massive, this theory shows up a mass gap.
The mass gap, that is also the mass of the lowest glue excitation, is given by 
\begin{equation}
\nonumber
      m_\sigma=
      \frac{\pi}{2K(i)}\sqrt{\sigma}=\frac{\pi}{2K(i)}(6\pi\alpha_s)^{\frac{1}{4}}\Lambda.
\end{equation}
%Decay is described by the following diagrams
%\begin{figure}[tbp]
%\begin{center}
%\includegraphics[width=50pt]{spipia.eps}
%\caption{\label{fig:fig1} $\sigma\rightarrow\pi^++\pi^-$ amplitudes.}
%\end{center}
%\end{figure}
Width is given by \cite{fra5} ($G'_{NJL}=\frac{3.76}{\sigma}g=3.76\frac{\sqrt{4\pi\alpha_s}}{\sigma}$)
\begin{equation}
\nonumber
   \Gamma_\sigma = \frac{2}{\pi}{G'}^{2}_{NJL}m_\sigma f_\pi^4\sqrt{1-\frac{4m_\pi^2}{m_\sigma^2}}
\end{equation}

Decay constants for all glue excitations can be straightforwardly obtained. Using the mapping theorem, one has a Fourier series for SU(3)
\begin{equation}
\scriptstyle{
   A_\mu^a(0,t)\approx\eta_\mu^a\sqrt{\frac{\sigma}{6\pi\alpha_s}}\frac{2\pi}{K(i)}\sum_{n=0}^\infty
   \sum_{n=0}^{\infty}(-1)^n
   \frac{e^{-\left(n+\frac{1}{2}\right)\pi}}{1+e^{-(2n+1)\pi}}\times
   e^{-im_nt-i\theta}+c.c.}
\end{equation}
From this series we can easily read the decay constants for the glue excitations
\begin{equation}
\nonumber
   f_{S_n}=\sqrt{\frac{\sigma}{6\pi\alpha_s}}\frac{2\pi}{K(i)}(-1)^n
   \frac{e^{-\left(n+\frac{1}{2}\right)\pi}}{1+e^{-(2n+1)\pi}}.
\end{equation}
Finally, for the $\sigma$, setting $n=0$, we have
\begin{equation}
\nonumber
   f_{\sigma}=\sqrt{\frac{\sigma}{6\pi\alpha_s}}\frac{2\pi}{K(i)}
   \frac{e^{-\frac{\pi}{2}}}{1+e^{-\pi}}.
\end{equation}
It is interesting to note that, using again the mapping theorem, the correlator is given by
\begin{eqnarray*}
   &&\langle A_\mu^a(0,t)A_\nu^b(0,0)\rangle = \eta_\mu^a\eta_\nu^b\langle\phi(0,t)\phi(0,0)\rangle= \\
   &&\sum_{n=0}^\infty B_ne^{-im_nt}e^{-i\theta}+c.c.+O\left(1/\sqrt{N}g\right)
\end{eqnarray*}
and so, eq.(\ref{eq:spe}) is indeed the spectrum of the theory in a strong coupling limit.

$\eta'$ decay is one of the key processes to understand glue role in hadronic physics. This is mainly $\eta'\rightarrow\eta\pi^+\pi^-$. Measures at DA$\Phi$NE proved that $\eta'$ has a significant glue component \cite{kloe}. So, we consider this decay as a two step process and essentially due to glue emission: $\eta'\rightarrow\eta\sigma$.
%\begin{figure}[tbp]
%\begin{center}
%\includegraphics[width=50pt, angle=90]{eta1.eps}
%\caption{\label{fig:fig1} $\sigma\rightarrow\pi^++\pi^-$ amplitudes.}
%\end{center}
%\end{figure}
This will give the following width
\begin{eqnarray*}
   \Gamma_{\eta'} &=& \frac{1}{2\pi}{G'}^{2}_{NJL}m_{\eta'}f_\sigma^2 f_\eta^2 \times \\
   &&\scriptstyle{\sqrt{\frac{m_{\eta'}^4+m_\eta^4+m_\sigma^4-2m_{\eta'}^2m_\eta^2
   -2m_{\eta'}^2m_\sigma^2-2m_\eta^2 m_\sigma^2}{m_{\eta'}^4}}}.
\end{eqnarray*}
The opposite process $\sigma\eta\rightarrow\eta'$ can be also easily computed by time reversal symmetry of QCD.
Glue production by $\eta'$ is a threshold process and so it can be used to fix $\sigma$ mass as
\begin{equation}
\nonumber
    m_\sigma\le m_{\eta'}-m_\eta.
\end{equation}
We can use the measured width \cite{pdg} to determine $f_\eta$. Taking for $\Gamma_{\eta'}$ the same value observed for the process $\eta'\rightarrow\eta\pi^+\pi^-$ and $f_\sigma$ the one computed above, one has for $m_\sigma$ at the threshold
\begin{equation}
\nonumber
   |f_\eta|\approx 0.019\ GeV.
\end{equation}
Now, we assume $\eta$ and $\eta'$ to mix with an angle $\theta\approx-14^\circ$ and also $f_\pi\approx 0.13\ GeV$, $f_0\approx -0.45f_\pi$ and $f_8\approx 1.2f_\pi$, so one has
\begin{equation}
\nonumber
   f_\eta=f_0\cos\theta+f_8\sin\theta\approx-0.019\ GeV.
\end{equation}
while $f_{\eta'}\approx 0.16\ GeV$. This computation does not depend on the value of $\alpha_s$.

\section{QCD and $\Lambda$ constant}

Now, we check above scenario against a proper value of $\alpha_s$. One has for the running coupling with six flavors\cite{pdg}
\begin{equation}
\nonumber
   \alpha_s(q^2,\Lambda)=\frac{1}{\frac{7}{4\pi}\ln\left(\frac{q^2}{\Lambda^2}\right)}
   \left(1-\frac{\frac{13}{8\pi^2}\ln\ln\left(\frac{q^2}{\Lambda^2}\right)}
   {\frac{7}{4\pi}\ln\left(\frac{q^2}{\Lambda^2}\right)}+\ldots\right)
\end{equation}
and we stop at this order to avoid dependencies from a renormalization scheme. Our aim is to fix $\Lambda$ in order to get a consistent scheme for low-energy QCD. This is the same parameter both for high and low-energy physics.
In our case, a key quantity is string tension. We will have
\begin{equation}
\nonumber
   \sigma(q^2,\Lambda)=\sqrt{6\pi\alpha_s(q^2,\Lambda)}\Lambda^2.
\end{equation}
So, the mass of the first glue state is obtained by solving the equation
\begin{equation}
\nonumber
    m_\sigma=\frac{\pi}{2K(i)}\sqrt{\sigma(m_\sigma^2,\Lambda)}.
\end{equation}
The next excited glue state will be given by 
\begin{equation}
\nonumber
    m_1=\frac{3\pi}{2K(i)}\sqrt{\sigma(m_1^2,\Lambda)}.
\end{equation}
We can write down the corresponding decay widths as
\begin{equation}
\nonumber
   \Gamma_\sigma = \frac{2}{\pi}{G'}^{2}_{NJL}(m_\sigma^2,\Lambda)m_\sigma f_\pi^4\sqrt{1-\frac{4m_\pi^2}{m_\sigma^2}}
\end{equation}
and for f0(980)
\begin{equation}
\nonumber
   \Gamma_{1\pi\pi} = \frac{2}{\pi}{G'}^{2}_{NJL}(m_1^2,\Lambda)m_1 f_\pi^4\sqrt{1-\frac{4m_\pi^2}{m_1^2}}.
\end{equation}
For the decay constants of glue states we will have
\begin{equation}
\nonumber
   f_{S_n}=\sqrt{\frac{\sigma(m_{S_n}^2,\Lambda)}{6\pi\alpha_s(m_{S_n}^2,\Lambda)}}\frac{2\pi}{K(i)}(-1)^n
   \frac{e^{-\left(n+\frac{1}{2}\right)\pi}}{1+e^{-(2n+1)\pi}}.
\end{equation}
On a similar ground we can write, as observed for f0(980) from its KK decay,
\begin{equation}
\nonumber
   \Gamma_{1KK} = \frac{2}{\pi}{G'}^{2}_{NJL}(m_1^2,\Lambda)m_1 f_K^4\sqrt{1-\frac{4m_K^2}{m_1^2}}.
\end{equation} 
We can compute the ratio $r_{f\pi K}=|g_{fKK}|/|g_{f\pi\pi}|$ predicted to be 2.59(1.34) by Mennessier , Narison, and Wang \cite{nar1}. These authors also agree with a content of f0(980) being mostly glue. We have
\begin{equation}
\nonumber
   g_{f\pi\pi} = 8\sqrt{6}m_f f_\pi^2\frac{\sqrt{4\pi\alpha_s}}{\sigma}
\end{equation}
and similarly for decay to Ks. This implies a sizable coupling of the $\sigma$ with K mesons.
This computation will represent a leading order approximation as we neglect mixing effects that should be anyhow present.
Finally, we will check our computation of $\Lambda$ against the corresponding value of $\alpha_s$ as obtained from other sources and given by PDG\cite{pdg}.

So our estimation is
\begin{equation}
\nonumber
   \Lambda=0.171\pm 0.001\ GeV
\end{equation}
that yields the following results for $\sigma$ mass
\begin{equation}
\nonumber
   m_\sigma=0.410\pm 0.007\ GeV,
\end{equation}
f0(980) mass
\begin{equation}
\nonumber
   m_{f_0(980)}=1.023\pm 0.002\ GeV,
\end{equation}
mass ratio
\begin{equation}
\nonumber
   m_{f0(980)}/m_\sigma= 2.49\pm 0.05,
\end{equation}
very near the theoretical value 3 but here we are considering running coupling and string tension. This may be relevant for understanding lattice results. $\sigma$ width is
\begin{equation}
\nonumber
   \Gamma_\sigma/2= 0.260\pm 0.001\ GeV.
\end{equation}
It is interesting to note that these values for width and mass are very near those of a recent analysis\cite{cap}.
Coupling ratio for f0(980)
\begin{equation}
\nonumber
   r_{f\pi K}= 1.42\pm 0.02
\end{equation}
in close agreement with Mennessier , Narison, and Wang \cite{nar1}. $\sigma$ decay constant is $f_\sigma= 0.139\pm 0.008\ GeV.$
%As a final check we have to verify the value of the running coupling that is $\alpha_s(M_Z^2,\Lambda)= 0.1277\pm 0.0001$ within $7\%$ error with respect to PDG best value \cite{pdg}.
Finally, we just note that the scenario given through this analysis matches rather well that given in a pioneering work of Narison and Veneziano\cite{nven,nar2}. With their choice of $\Lambda$, they get a width for $\sigma$ very near to the correct one obtainable from our formulas and a corresponding decay constant increasing with mass. 
\nin
%%%%%%%%%%%%%%%%
\section{Conclusions}
\nin
%%%%%%%%%%%%%%%%
Low-energy limit of QCD can be obtained by an expansion at very large coupling and remapping Yang-Mills field on a quartic massless scalar field theory. This mapping is a proved mathematical theorem. A mass gap arises due to strong self-interaction of the fields already at classical level.
QCD infrared limit is a renormalizable Yukawa model that, in the proper approximation, reduces to a Nambu-Jona-Lasinio model. This is possible as we have an analytical formula, in closed form, for the gluon propagator in close agreement with lattice data. A lot of hadronic processes can be described by a generating functional obtained in closed form from QCD in the above limit but neglecting quark loops and quark-quark interaction and having only quark-glue interactions. From this, one can compute decay constants, width and several other experimental meaningful observables. $\sigma$ and f0(980) appear to be glue excitations entering into a lot of hadronic processes and describing the true physical states of glue for low-energy QCD. It is worthwhile to note that decay constants for glue excitations are strongly damped for higher excitations making observable just the first few states. $\eta'$ appears mostly to emit a glue state that, due to a threshold condition, should properly fix the value of the mass of $\sigma$ resonance. In agreement with KLOE-2 evidences, this gives a sound explanation of the main decay of this meson resembling the case of QED of an excited atom decaying into an atom plus the force carrier, i.e. a photon. This decay can also be used to determine the decay constant of $\eta$. All observables are properly fixed with a single constant $\Lambda=171\pm 1\ MeV$ that recovers the correct value of the measured running coupling at $M_Z$.
%%%%%%%%%%%%%%%%%%%%%%%%%%%
%\section*{Acknowledgements}
%\nin
%%%%%%%%%%%%%%%%



%% References


%\bibliographystyle{elsarticle-num}


%%%%%%%%%%%%%%%%%%%%
%\vfill\eject

%\input{bib_sample}

\end{document}